\documentclass{moriond}
\usepackage{amsmath}
\usepackage{amssymb}

\def\Journal#1#2#3#4{{#1} {\bf #2}, #3 (#4)}
\def\AA{\em Astron. Astrophys.}

\def\PRL{\em Phys. Rev. Lett.}
\def\PRD{{\em Phys. Rev.} D}
\def\JCAP{\em J. Cosmology Astropart. Phys.}
\def\MNRAS{\em Mon. Not. R. Astron. Soc.}

\def\be{\begin{equation}}
\def\ee{\end{equation}}
\def\bea{\begin{eqnarray}}
\def\eea{\end{eqnarray}}

\newcommand{\dzsz}{{\tt d0s0}}
\newcommand{\doso}{{\tt d1s1}}
\newcommand{\dmsm}{{\tt dmsm}}
\newcommand{\dtsf}{{\tt d10s5}}

\begin{document}
\vspace*{2cm}
\title{The Simons Observatory: component separation pipelines for $B$-modes}

\author{K. Wolz$^1$, S. Azzoni$^{2}$, C. Herv\'{\i}as-Caimapo$^{3}$, J. Errard$^{4}$, N. Krachmalnicoff$^{5,6,7}$, D. Alonso$^{1}$, \\B. Beringue$^{4}$, E. Hertig$^{8}$, on behalf of the Simons Observatory Collaboration\vspace*{4mm}}

\address{$^1$ Department of Physics, Keble Road, Oxford OX1 3RH, UK\vspace*{-2mm}}
\address{$^2$ Joseph Henry Laboratories of Physics, Jadwin Hall, Princeton University, Princeton, NJ 08544, USA\vspace*{-2mm}}
\address{$^3$ Instituto de Astrof\'isica and Centro de Astro-Ingenier\'ia, Facultad de F\'isica, Pontificia Universidad Cat\'olica de Chile, Av. Vicu\~na Mackenna 4860, 7820436 Macul, Santiago, Chile\vspace*{-2mm}}
\address{$^4$ Universit\'{e} Paris Cit\'e, CNRS, Astroparticule et Cosmologie, F-75013 Paris, France\vspace*{-2mm}}
\address{$^5$ International School for Advanced Studies (SISSA), Via Bonomea 265, I-34136 Trieste, Italy\vspace*{-2mm}}
\address{$^6$ National Institute for Nuclear Physics (INFN) -- Sezione di Trieste, Via Valerio 2, I-34127 Trieste, Italy\vspace*{-2mm}}
\address{$^7$ Institute for Fundamental Physics of the Universe (IFPU), Via Beirut 2, I-34151 Grignano (TS), Italy\vspace*{-2mm}}
\address{$^8$ IoA and KICC, University of Cambridge, Madingley Road, Cambridge, CB3 0HA, UK}

\maketitle\abstracts{The upcoming Simons Observatory (SO) Small Aperture Telescopes aim at observing the degree-scale anisotropies of the polarized CMB to constrain the primordial tensor-to-scalar ratio $r$ at the level of $\sigma(r=0)\lesssim0.003$ to probe models of the very early Universe. We present three complementary $r$ inference pipelines and compare their results on a set of sky simulations that allow us to explore a number of Galactic foreground and instrumental noise models, relevant for SO. In most scenarios, the pipelines retrieve consistent and unbiased results. However, several complex foreground scenarios lead to a $>2\sigma$ bias on $r$ if analyzed with the default versions of these pipelines, highlighting the need for more sophisticated pipeline components that marginalize over foreground residuals. We present two such extensions, using power-spectrum-based and map-based methods, and show that they fully reduce the bias on $r$ to sub-sigma level in all scenarios, and at a moderate cost in terms of $\sigma(r)$.}

\section{Introduction}

  One of the next frontiers in cosmology is the potential detection of primordial gravitational waves through the large-scale $B$-modes in the cosmic microwave background (CMB). Primordial tensor perturbations, if existing, would form a stochastic background of gravitational waves, sourcing a parity-odd $B$-mode component in the polarization of the CMB. \cite{mk97} The amplitude of these tensor perturbations is parameterized by the tensor-to-scalar ratio $r$. An unequivocal measurement of $r$, or a stringent upper bound, would greatly constrain the landscape of theories of the early Universe. Current CMB experiments find $r < 0.032$ at 95\% CL.\cite{mt21} 

  Several factors complicate measuring $r$: first, the gravitational deflection of the background CMB photons by the cosmic large-scale structure creates coherent sub-degree distortions in the CMB, known as CMB lensing, transforming a fraction of the parity-even $E$-modes into $B$-modes \cite{zs98}. Second, diffuse Galactic foreground emission, mostly thermal dust and synchrotron radiation, sources $B$-modes at a level which would exceed any primordial signal. Component separation methods, which exploit the different spectral energy distributions (SED) to separate the CMB from foregrounds, are thus of vital importance. On the other hand, using overly simple models to describe the potentially complex emission in real foregrounds, may lead to systematic residuals and significantly bias an $r\sim10^{-3}$ measurement.\cite{nk18} Component separation pipelines must therefore be robust against instrumental noise and systematic effects.
  
  The Simons Observatory (SO) targets the detection of an $r\sim0.01$ signal.\cite{so19} SO is a ground-based experiment, located at the Cerro Toco site in the Chilean Atacama desert, which observes the microwave sky in six frequency channels, from 27 to 280\,GHz, with full science observations scheduled to start in 2024. SO consists of a Large Aperture Telescope (LAT) with a 6m diameter aperture that is key to subtracting lensing-induced $B$-modes using a technique  known as ``delensing''.\cite{tn22} On the other hand, multiple Small Aperture Telescopes (SATs) with 0.4m diameter apertures will make deep larger-scale maps of $\sim 10$\% of the sky, targeting primordial $B$-modes.

  In this work, based on Wolz {\it et al.} (2023),\cite{so23} we validate three independent $B$-mode analysis pipelines, comparing their performance at a potential $r$ detection by the SO SATs, and evaluating the capability of the survey to constrain $\sigma(r=0) \leq 0.003$ in the presence of foregrounds and instrumental noise. To that end, we produce SAT-like sky simulations with realistic foregrounds of varying complexity, primordial and lensing $B$-modes, and instrumental noise,\footnote{We anticipate a detailed study of filtering-related and instrumental systematics in a future work.} computed from parametric models in the SO science goals and forecast paper (hereafter SO19).\cite{so19}

\section{Pipelines}
  Our three component separation pipelines adopt complementary techniques widely used in the literature: power-spectrum-based parametric cleaning (pipeline A), Needlet Internal Linear Combination (NILC) blind cleaning (pipeline B), and map-based parametric cleaning (C).

  \textit{Pipeline A} uses multi-frequency power-spectrum-based component separation, similar to the BICEP/{\sl Keck} method.\cite{bk18} The data vector contains the $BB$ cross-frequency power spectra by the \texttt{NaMaster} package.\cite{da19} We average over cross-split spectra to avoid noise bias, and use $B$-mode purification to minimize the $E$-to-$B$ polarization leakage caused by masking the sky. We consider a Gaussian likelihood with a data model that contains nine free parameters describing CMB $B$-modes ($r$, $A_{\rm lens}$), spatially isotropic Galactic dust and synchrotron emission. The covariance is estimated from 500 simulated sky maps with coadded Gaussian CMB, dust, synchrotron and inhomogeneous noise. Spatially varying spectral indices give rise to frequency decorrelation, which we can optionally model by the moment expansion, introducing four extra parameters.\cite{sa21} The full pipeline is publicly available.\footnote{See \href{https://github.com/simonsobs/BBPower}{github.com/simonsobs/BBPower}.}

  \textit{Pipeline B} is based on the blind Internal Linear Combination (ILC) method, which removes any contaminants (foregrounds, noise) that do not possess a blackbody distribution like the CMB.\cite{sb13} The method analyzes data in needlet space, allowing to capture features localized in both real and harmonic space, and reconstructs a clean CMB map from a linear combination of the frequency maps. We then compute the $BB$ power spectrum of this map using \texttt{NaMaster}, and subtract the noise bias obtained from noise simulations. We then infer $r$ and $A_{\rm lens}$ using a Gaussian power spectrum likelihood with an empirical covariance obtained from 500 sky simulations.

  \textit{Pipeline C} is a map-based parametric pipeline based on the \texttt{fgbuster} code \cite{fg23}. This method reconstructs sky components by finding linear combinations of frequency maps by maximizing a spectral likelihood that assumes parametric models for the (isotropic) spectral energy distributions of dust and synchrotron. After obtaining the cleaned CMB map, we proceed analogously to pipeline B. As a possible answer to frequency decorrelation, we can extend pipeline C by marginalizing over a residual dust amplitude parameter in the final power spectrum likelihood.

\section{Simulations}

  We built a set of dedicated simulations against which to test our data analysis pipelines and compare results. The simulated maps include the cosmological CMB signal, Galactic foreground emission as well as instrumental noise. We model the SO-SAT noise power spectra as a sum of white detector noise (considering a ``baseline'' and a ``goal'' level) and filtering-related $1/f$-noise (considering a ``pessimistic'' and an ``optimistic'' case), introduced in SO19. We draw 500 Gaussian map realizations and modulate them with the SO19 SAT inverse hit counts to mimic the survey anisotropy. We draw 500 Gaussian random realizations of the CMB following the \textit{Planck} 2018 cosmology. Our default input model has only lensing $B$-modes ($r=0\, , \; A_{\rm lens}=1$), but we also consider nonzero primordial $B$-modes ($r=0.01$) and 50\% delensing ($A_{\rm lens}=0.5$), achievable for SO.\cite{tn22} For the six frequency channels, we convolve signal simulations with Gaussian beams of FWHMs of (91, 63, 30, 17, 11, 9) arcmin, respectively. Galactic thermal dust and synchrotron emission are known to be the main contaminants to CMB observations in polarization. Many aspects of their emission remain unconstrained, especially the characterization of the variation of their spectral energy distribution (SED) across the sky. To assess their impact on $r$ inference, we use five different models of Galactic synchrotron and dust emission with increasing levels of complexity, based on the {\tt PySM} package \cite{bt17}: Gaussian (isotropic), \dzsz{} (anisotropic amplitude, isotropic SEDs), \doso, \dmsm, and \dtsf{} (anisotropic, with progressively larger SED variations across the sky).

\section{Results and discussion}

\begin{figure}
    \centering
    \includegraphics[width=0.52\columnwidth]{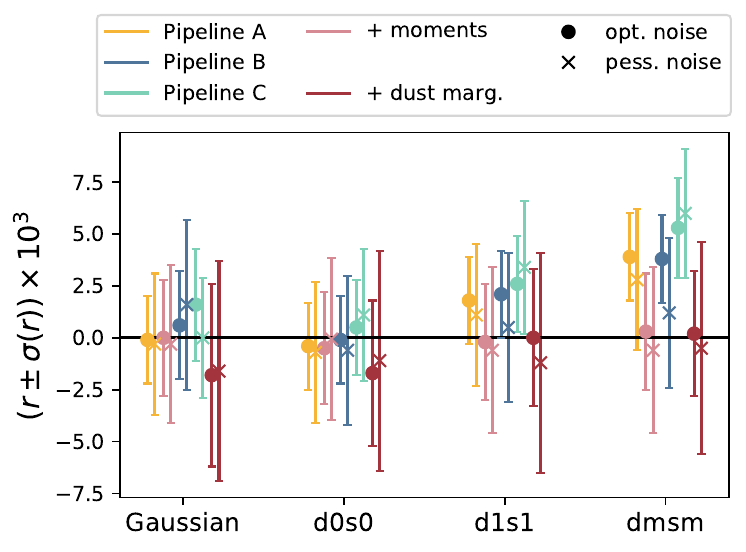}
    \includegraphics[width=0.46\columnwidth]{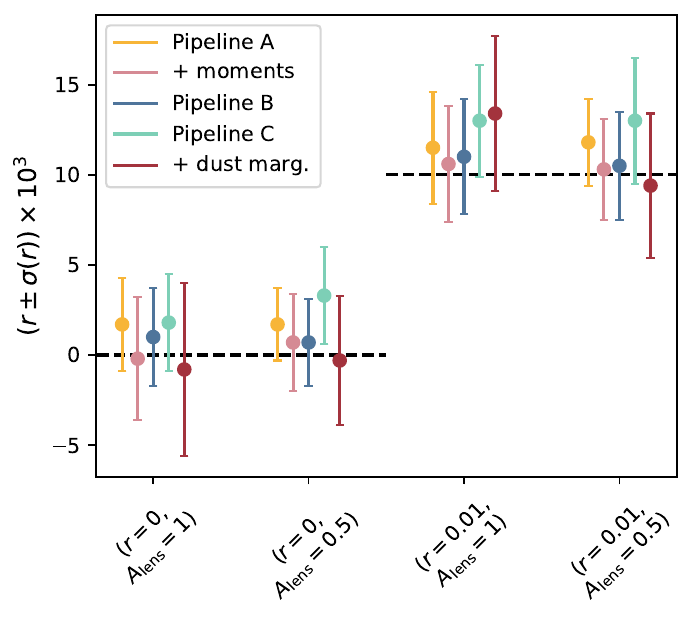}
    \caption{Mean $r$ with 68\% CL after applying the nominal component separation pipelines (plus extensions) to 500 simulations. \emph{Left:} comparison of foregrounds with increasing complexity, assuming a fiducial cosmology with $r=0$ and $A_{\rm lens}=1$, inhomogeneous noise with goal sensitivity and optimistic $1/f$ noise component (dot markers), and inhomogeneous noise with baseline sensitivity and pessimistic $1/f$ noise component (cross markers). \emph{Right:} comparison of input models including primordial $B$-modes and 50\% delensing efficiency, the SO baseline noise level with optimistic $1/f$ component, and the \doso{} foreground template.}
    \label{fig:results}
\end{figure}

  Figure~\ref{fig:results} on the left shows the mean $r$ and 68\% credible intervals found by each pipeline as a function of the input foreground model. We show five setups: pipeline A nominal, pipeline A + moments, pipeline B, pipeline C, and pipeline C + dust marginalization. Dots or crosses denote goal white noise with optimistic $1/f$ noise, or baseline white noise with pessimistic $1/f$ noise, respectively. As expected, the nominal pipelines are unbiased for the Gaussian and \dzsz{} foregrounds, while \doso{} and \dmsm{} cause a bias of up to $2\sigma$. This is plausible, since the progressively larger SED anisotropy is expected to lead to an increased bias on $r$ if unaccounted in the model. All pipelines achieve comparable statistical uncertainty on $r$, ranging from $\sigma(10^3 r)=2.1$ to 3.6 depending on the noise model. Changing between goal and baseline white noise level results in an increase of $\sigma(r)$ of $\sim 20$-$30\%$. Changing between optimistic and pessimistic $1/f$ noise has a similar effect for pipelines A and B, and somewhat weaker for pipeline C. These results are compatible with the forecasts presented in SO19. The extended pipelines A and C are able to reduce the bias on $r$ to below $1\sigma$ in all noise and foreground scenarios. For the Gaussian and \dzsz{} foregrounds, we consistently observe a small negative bias (of below $0.5\sigma$), potentially caused by volume effects from extra parameters that are poorly constrained by the data (to be studied in a future work). For \doso{}{} and \dmsm{}, both pipeline extensions fully reduce the bias to below $0.5\sigma$. Depending on the input noise case, using the extended pipeline increases the average statistical uncertainty by $\sim$25\% for pipeline A, and by $\sim$80\% for pipeline C.

  We repeated this analysis for input cosmologies with $r\in\{0,\, 0.01\},\, A_{\rm lens}\in \{0.5,\, 1\}$. For simplicity, we only considered baseline white noise with optimistic $1/f$ noise in the \doso{} foreground scenario. As seen in Fig.~\ref{fig:results} on the right, a 50\% delensing efficiency reduces $\sigma(r)$ by up to 30\%.\footnote{A related work found 37\% improvement on $\sigma(r)$ using template-based delensing.\cite{eh24}} The presence of primordial $B$-modes of $r=0.01$ results in larger cosmic variance, leading to a $\sim 40\%$ increase in $\sigma(r)$, in agreement with theoretical expectations. For C + dust marginalization without delensing, we even see a decrease in $\sigma(r)$, hinting at a possible degeneracy between $r$ and the dust amplitude. Concluding, all pipelines are able to detect the $r=0.01$ signal at a level of $\sim3\sigma$, and both pipeline extensions eliminate the observed 0.5-1.2$\sigma$ bias on $r$.

  While completing this work, the new {\tt PySM} Galactic emission models were published,\footnote{See \href{https://pysm3.readthedocs.io/en/latest}{pysm3.readthedocs.io/en/latest}.} including the preliminary model \dtsf{}, which we added to our analysis. The marginalized posterior mean and 68\% credible intervals on $10^3 r$, averaged over 100 simulations, are $19.2\pm1.9$, $15.8\pm1.9$, and $22.0\pm 2.6$ for the nominal pipelines A, B, and C, respectively, corresponding to biases of 10, 8, and 8$\sigma$. Fortunately, with both extensions of pipelines A and C this reduces to sub-sigma biases, with $10^3 r = 2.8\pm2.8$ and $-1.5\pm5.1$, respectively. These results consider the most optimistic noise scenario, while other cases would likely lead to lower relative biases. Overall, it is encouraging that two pipeline extensions are able to robustly separate the cosmological signal from complex models such as \dtsf. This highlights the importance of marginalizing over foreground residuals given the level of sensitivity achievable by SO.

  Concluding, this work, based on Wolz {\it et al.} (2023),\cite{so23} shows that the pipelines in place for SO are able to robustly constrain the primordial tensor-to-scalar ratio in the presence of Galactic foregrounds covering the full range of complexity envisaged by current, state-of-the-art models.

\section*{Acknowledgments}

  This work was supported in part by a grant from the Simons Foundation (Award 457687, B.K.).

\end{document}